\documentclass[11pt,prd,onecolumn,showpacs,amsmath,amssymb,aps,floats,floatfix,nofootinbib]{revtex4-1}
\usepackage[colorlinks=true,urlcolor=rossos,anchorcolor=black,citecolor=mygreen,filecolor=black,linkcolor=black,menucolor=blue,pagecolor=blue,linktocpage=true]{hyperref} 
\usepackage[multidot]{grffile}  
\usepackage{dcolumn}
\usepackage{bm}
\usepackage{amsmath}
\usepackage{amsfonts}
\usepackage{amssymb}
\usepackage{color}
\usepackage{footmisc}
\usepackage{latexsym}
\usepackage{slashed} 
\usepackage{pstricks}
\usepackage{indentfirst}
\usepackage{mathrsfs}
\usepackage{multirow}
\usepackage{epsfig,psfrag}
\usepackage{subfigure}
\usepackage{mathtools}
\usepackage{enumitem}
\usepackage{setspace} 
\usepackage[utf8]{inputenc} 
\usepackage[scientific-notation=true]{siunitx} 
\graphicspath{{figs/}}

\newcommand{\R}{\mathcal{R}}

\definecolor{rossos}{cmyk}{0,1,1,0.55}
\definecolor{mygreen}{rgb}{0.27, 0.64, 0.48}

\makeatother

\allowdisplaybreaks 

\begin{document}

\title{\Large \bf \color{rossos} Detecting Axion Dark Matter through the Radio Signal \\from Omega Centauri}

\author{\bf Jin-Wei Wang$^{a,b,c}$}
\email{\color{black}jinwei.wang@sissa.it}
\author{Xiao-Jun Bi$^{d,e}$}
\email{\color{black}bixj@ihep.ac.cn}
\author{Peng-Fei Yin$^{d}$}
\email{\color{black}yinpf@ihep.ac.cn}

\affiliation{\vspace{0.4cm}$^a$Scuola Internazionale Superiore di Studi Avanzati (SISSA), via Bonomea 265, 34136 Trieste, Italy}
\affiliation{$^b$INFN, Sezione di Trieste, via Valerio 2, 34127 Trieste, Italy}
\affiliation{$^c$ Institute for Fundamental Physics of the Universe (IFPU), via Beirut 2, 34151 Trieste, Italy}
\affiliation{$^d$Key Laboratory of Particle Astrophysics, Institute of High Energy Physics, Chinese Academy of Sciences, Beijing, China}
\affiliation{$^e$School of Physical Sciences, University of Chinese Academy of Sciences, Beijing, China}

\begin{abstract}
\large
\vspace{0.4cm}
As a well-motivated dark matter candidate, axions can be detected through the axion-photon resonant conversion in the magnetospheres of magnetic white dwarf stars or neutron stars. In this work, we utilize Omega Centauri, which is the largest globular cluster in the Milky Way and is suggested to be the remnant core of a dwarf galaxy, to probe the axion dark matter through radio signals that originate from all the neutron stars and magnetic white dwarf stars in it.
With 100 hours of observation, the combination of SKA phase 1 and LOFAR can effectively probe the parameter space of the axion-photon coupling $g_{a\gamma}$ up to $10^{-14}\sim 10^{-15}~\text{GeV}^{-1}$ for the axion mass range of $0.1\sim 30 ~\mu\text{eV}$. Depending on the choice of neutron star evolution model, this limitation is two or three and a half orders of magnitude higher than that of the single neutron star or magnetic white dwarf.
\end{abstract}

\maketitle
\tableofcontents

\section{Introduction}
\label{sec:intro}

The quantum chromodynamics (QCD) axion, which was proposed to solve the strong CP problem \cite{PhysRevD.16.1791,PhysRevLett.38.1440,PhysRevLett.40.223,PhysRevLett.40.279}, has been regarded as one of the most compelling dark matter (DM) candidates and gained more and more attention in the past decade \cite{PRESKILL1983127,ABBOTT1983133,DINE1983137,1510.07633,1602.00039,2003.01100}.
Based on the coupling between axion and electromagnetic sector, i.e.  $\mathcal{L}_{a\gamma\gamma} = -(1/4) g_{a\gamma} a F_{\mu\nu} \tilde{F}^{\mu\nu} = g_{a\gamma} a \bm{E} \cdot \bm{B}$, it suggests that the axion DM could be detected through the axion-photon conversion process, which can happen in the presence of external magnetic field. As the axion-photon conversion probability is positively associated with the magnetic field strength \cite{1804.03145,2101.02585}, the compact stars, (CSs\footnote{In this paper, the CSs always stands for NSs and/or MWDs unless specifically stated.}), including magnetic white dwarf stars (MWDs) and neutron stars (NSs),
are potentially very promising probes to search for the axion DM because of their extremely strong magnetic
fields. Besides, due to the existence of plasma in CSs' magnetosphere, the conversion probability can be further greatly enhanced (called resonant conversion) when the axion mass $m_a$
is equal to the plasma frequency $\omega_p$ \cite{1804.03145,2101.02585}. Interestingly, the plasma frequency of a typical CS happens to be in the radio frequency band \cite{1804.03145,2101.02585}. This means that the axion-induced signal could be detected by radio telescopes, such as the Square Kilometer Array (SKA) \cite{skainfo,skadesign,skapro} and the LOw-Frequency ARray
(LOFAR) \cite{1305.3550}.

Recently, there have been several investigations utilizing the single isolate  NS \cite{0711.1264,1803.08230,1804.03145,1912.08815,2008.01877,Prabhu:2021zve,2104.07670,2104.08290,Buckley:2020fmh,Millar:2021gzs} or MWD \cite{2101.02585} to detect axion DM through radio signal that originates from the axion-photon conversion in magnetosphere of these CSs. The results show that both targets give comparable sensitivities, that is, the lower limit of $g_{a\gamma}$ from the SKA phase 1 (SKA1) with 100 hours of observation can reach  $\sim 10^{-12}$ GeV$^{-1}$ for the $\sim \mu$eV axion \cite{2101.02585}.
Based on these works, it could be also intriguing to investigate the signature of axion DM from the astrophysical systems that contains a lot of CSs, such as globular clusters (GCs) \cite{1811.01020}. It can be expected to get a much stronger signal from these systems than a single CS.

In this work, we propose to use the GC Omega Centauri ($\omega$ Cen) as a probe to detect the axion DM. Among the Milky Way's $\sim$ 200 GCs, the $\omega$ Cen is very unique. It is the most massive ($\sim 4.05\times10^6 M_\odot$), the most luminous, and has the largest core and half-light radius \cite{stw2488,Harris:1996kt,1907.08564,1211.4399}. Besides, $\omega$ Cen also possesses multiple stellar populations with a large spread in metallicity 	and spatial distributions that include a trailing stellar stream \cite{Piotto:2004mf,Sollima:2005qy,natureLee,2019NatAs}. All these observations suggest that $\omega$ Cen may be the remnant core of a dwarf galaxy whose outskirts were tidally stripped as it fell into the Milky Way \cite{1907.08564,2019NatAs,Bekki2003}. If so, that means the $\omega$ Cen probably contains amounts of DM and thus has a much higher DM density compared to the local DM density $\sim0.3 ~\text{GeV}/\text{cm}^{3}$ (see Sec. \ref{sec:31})\footnote{Note that whether or not GCs are born within DM mini-halos is still under debate. For example, the observation of GC NGC 3201 shows that there is no direct evidence for the existence of DM in it \cite{Bianchini_2019,Wan_2021}, while for GC 47 Tuc the maximum likelihood analysis reveals that DM component is significantly preferred ($\sim 1\%$ total mass of GC 47 Tuc) \cite{Brown:2018pwq}. The masses of these two GCs are about one order of magnitude smaller than $\omega$ Cen. The analysis in Ref. \cite{1907.08564} suggests that there are amounts of DM in $\omega$ Cen (see Sec. \ref{sec:31}).}.
In addition, astronomical observations indicate that the DM velocity dispersion in $\omega$ Cen is around $\sim$ 30 km/s, which is about one order of magnitude smaller than that in the Milky Way field. As shown in Ref. \cite{1804.03145}, a smaller velocity dispersion can render a larger energy flux density.
Finally, as a rough estimate, the stellar mass in $\omega$ Cen is about $\sim$ $3\times10^6 M_\odot$. This means that there are about $\sim 10^4$ NSs and $\sim 10^5$ MWDs (see Sec. \ref{sec:32}). Consequently, the combination of so many "isolate sources" can greatly enhance the observed signal strength.
Therefore, all of these unique properties/advantages make the $\omega$ Cen an ideal detection target compared with a single NS or MWD \cite{2101.02585}.

This paper is organized as follows. In Sec. \ref{sec:starcomp} we introduce the properties of individual or ensemble of CSs, including the structure of the magnetic field and the plasma, the evolution and/or distribution of magnetic field strength, mass, and spin period. In Sec. \ref{sec:starDM} we introduce  the properties of $\omega$ Cen, e.g. the DM profile, the number and spatial distribution of CSs, and characteristics of the frequency spectrum. In Sec. \ref{sec:telescope} we estimate the sensitivity of SKA1 and LOFAR. In Sec. \ref{sec:res} we calculate the radio flux density of $\omega$ Cen as well as the constrains on the axion-photon coupling strength $g_{a\gamma}$ at the SKA1 and LOFAR. Conclusions and discussions are given in Sec. \ref{sec:conclusion}.

\section{The properties of compact stars}
\label{sec:starcomp}
In order to calculate the axion-induced radio signal from the ensemble of CSs, it is necessary to know some specific parameters of CSs, such as the mass, radius, magnetic field strength and so on.
In this section, we construct the population models of CSs to describe  the distribution and/or evolution of these important parameters. Note that some of the content has already been covered in previous studies \cite{1804.03145,2101.02585,1811.01020}, so here we just show the necessary conclusions for this part.

\subsection{The configuration of magnetic field and plasma}
\label{sec:Bstructure}

For the magnetic field of a single CS, we adopt the dipole configuration and assume that its rotation axis is parallel or anti-parallel to the magnetization axis\footnote{This is actually a reasonable assumption. For NSs, this misalignment angle tends to 0 (or $\pi$) as the evolution of NSs, although its initial value could be an order one number (see \ref{sec:NSproperty}). For MWDs, we have adopted the static plasma distribution assumptions (see below), which are independent of magnetic field, so for an ensemble of MWDs, the effect of this misalignment angle can be absorbed into the angle between line-of-sight and rotation axis.} \cite{1903.05088,2101.02585}:
\begin{align}\label{eq:B_dipole}
	\bm B = \frac{B_0}{2} \, \frac{\R^3}{r^3} \, \Bigl( 3 (\hat{\bm m} \cdot \hat{\bm r}) \, \hat{\bm r} - \hat{\bm m} \Bigr)
	\qquad \text{for} \quad r > \R, \,
\end{align}
where $\R$ is the radius of CSs, $B_0$ is the value of the magnetic field strength at the CSs' surface in the direction of the magnetic pole,  $\bm{m} = 2\pi B_0 \R^3 \hat{\bm m}$ is the magnetic dipole moment, $\bm r = r \hat{\bm r}$ is the spatial coordinate, and $r = |\bm r|$ represents the distance from the center of CSs. Clearly, we can find that in this case the direction of $\bm B$ only depends on the $\theta$, which denotes the angle between $\hat{\bm m}$ and $\hat{\bm r}$. Its magnitude can be expressed as \cite{1811.01020,2101.02585}:
\begin{align}\label{eq:B_dipole2}
	B= |\bm B| =\frac{B_0}{2} \, \frac{\R^3}{r^3} \sqrt{3 ~\text{cos}^2\theta+1} \qquad \text{for} \quad r > \R. \,
\end{align}

In addition to the magnetic field structure of the CSs, the plasma distribution is also crucial to the calculation of the axion-photon conversion probability.
For the structure properties of the MWDs' coronae, we adopt the same assumptions as in Ref. \cite{V.Zheleznyakov,2101.02585}: (1) the corona is composed of fully ionized hydrogen plasma uniformly covering the entire surface of MWDs; (2) the field-aligned temperature of the electrons $T_\text{cor}$ is a constant throughout the corona. Under these conditions the distribution of the electron density is described by the barometric formula \cite{V.Zheleznyakov,Weisskopf_2007}.

The situation for NSs, however, is more complicated. As pointed in Ref. \cite{1811.01020}, the active NSs only makes up a very small fraction ($\sim 0.4\%$) of the total NS population, while the rest are dead NSs. With the evolution of NSs, the rotation period $P$ becomes larger and larger. At some point NSs will not be able to sustain the voltage required for the pair production. This happens around the so-called "death line" (see Ref. \cite{astro-ph/0001341} for details). For the evolution of NSs, there are two different models, which are dubbed model 1 and model 2 (see Sec. \ref{sec:NSproperty}) \cite{FaucherGiguere:2005ny,2010MNRAS.401.2675P}. The key difference between these two models is that the model 2 allows the NSs' magnetic field strength to decay with time, while in model 1 the magnetic field does not vary with time.
Note that in this work, we adopt the same approximate criteria for both models, that is to say, the active NSs fulfill $B_0/P^2 > 0.34\times10^{12} \text{~G s}^{-2}$ \cite{Bhattacharya1992}.

For the active NSs, the plasma charge density is given by the Goldreich-Julian
model \cite{1969ApJ,1811.01020}, while for the dead NSs the electron density profile is described by electrosphere model, which is derived by solving the time-dependent plasma dynamics with Maxwell equations on a spherical grid \cite{1811.01020}. Interestingly, as verified in Ref. \cite{1811.01020}, for an ensemble of NSs, the plasma density profile is a sub-dominant source of uncertainty compared to the uncertainty between e.g. NS model 1 and model 2 themselves. Therefore, we simply ignore the differences between the two plasma density distribution models and use the Goldreich-Julian model to describe both active and dead NSs. More technical details can be found in Ref. \cite{1811.01020} and the relevant references therein.

Once we have the plasma density profile of CSs, the resonant conversion radius $r_c$ can be derived by using the resonant conversion condition, i.e. $m_a = \omega_p$. The resonant radius of MWDs and NSs are given by \cite{1804.03145,2101.02585}
\begin{align}
	\label{resrWD}
	r_{c}^\text{WD} & = R_\text{WD} + 21.90 \times \left[2.634+\text{ln}\left(\frac{n_{e0}}{10^{10} ~\text{cm}^{-3}}\right) + \text{ln}\left(\frac{\mu \text{eV}^2}{m_a^2}\right)\right]\notag\\
	&\qquad\qquad\qquad~~~  \times\left(\frac{T_\text{cor}}{10^6 ~\text{K}}\right) \left(\frac{M_\text{WD}}{M_\odot}\right)\left(\frac{R_\text{WD}}{10^4 ~\text{km}}\right)^{-2} \text{km},
\end{align}
\begin{equation}
	\label{resrNS}
	r_c^\text{NS}   = 168.62 \, \, \times \big|3 \cos^2 \theta  - 1 \big|^{1/3} \times
	\left( {R_\text{NS} \over 10 \, \, \text{km}} \right) \times \left[  {B_0 \over 10^{14} \, \, {\rm G} } \ { 1\, \, \text{sec} \over P} \left( {1 \, \, \mu \text{eV} \over m_a} \right)^2 \right]^{1/3} \text{km} \,,
\end{equation}
where $n_{e0}$ is the electron density at the base of the MWDs' corona, $T_\text{cor}$ is the temperature of MWDs' corona, $R_\text{WD}$ ($R_\text{NS}$) is the radius of MWDs (NSs), $M_\text{WD}$ is the mass of the MWDs, and $P$ is the rotation period of NSs.

\subsection{The properties of magnetic white dwarf stars}
\label{sec:WDproperty}

For each individual MWD, there are five parameters need to be fixed in order to calculate the corresponding $r_{c}^\text{WD}$ and energy flux density $S_{a\gamma}^\text{WD}$ (see Eq. \eqref{resrWD} and \eqref{eq:WDsingal}). Thereinto, the $n_{e0}$  and $T_\text{cor}$ closely depend on the nature of the MWD's corona. For simplicity, we set $n_{e0} = 10^{10} ~\text{cm}^{-3}$ and $T_\text{cor} = 10^6 ~\text{K}$ as benchmark, which fulfill the constraints set by X-radiation searches \cite{V.Zheleznyakov,Weisskopf_2007,2101.02585}.
Besides, the mass and radius of a MWD are not independent variables, and they are related by the equation of state.
Therefore, there are only two parameters, i.e. $R_\text{WD}$ and $B_0$, need to be determined for each individual MWD.
\begin{figure}	
	\centering
	$$\includegraphics[width=0.47\textwidth]{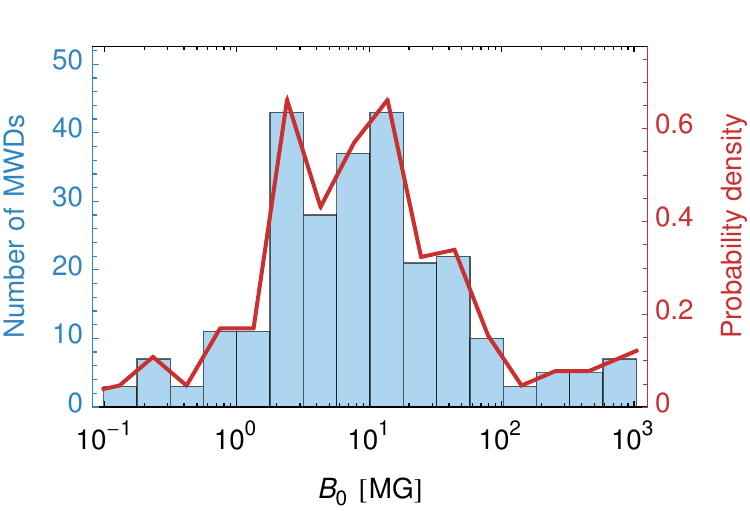}~~~~
	\includegraphics[width=0.47\textwidth]{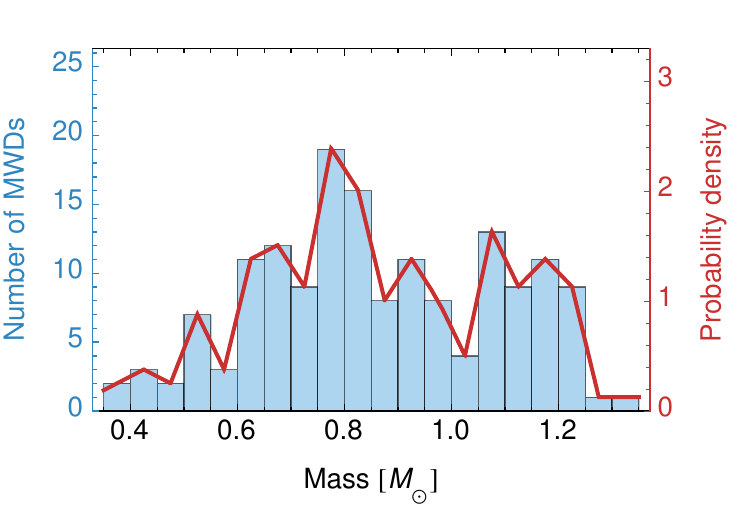}
	$$
	\vspace{-1cm}
	\caption{\em
		On the left: the magnetic field distribution of isolated MWDs \cite{2020AdSpR}. The red solid line represents the approximate PDF with linear interpolation. On the right: same plot as the left one, but for the mass of MWDs \cite{2020AdSpR}.}
	\label{figs1}
\end{figure}

However, the origin of the MWDs' magnetic field and the incidence of magnetism in WDs remains poorly estimated \cite{Hollands_2015,stw1357}. The volume limited samples of nearby WDs present the most unbiased estimates of the magnetic incidence when considering all WD sub types. It suggests a $13\pm4\%$ incidence of magnetism for WD sample within 20 pc \cite{Kawka_2007}, and this result is further verified by the work of Holberg et al \cite{stw1357}. According to the result in Ref. \cite{Kawka_2007}, the portion of MWDs with $B_0\gtrsim0.1 ~\text{MG}$ is about $\sim9.5\%$ among all WDs.

Instead of a single MWD, in order to calculate the radio signal of an ensemble of MWDs, it is necessary to know their magnetic field strength and mass distributions, which are summarized in a recent review of MWDs (see Fig. \ref{figs1}) \cite{2020AdSpR}.
The blue bars represent the magnetic field strength (left) and mass (right) statistical distribution of the MWDs. These statistics include hundreds of samples of MWDs. The red solid lines are the approximate probability density functions (PDFs) that are derived by using the linear fitting.
In this work, we assume that these PDFs represent the real mass and magnetic field distributions of MWDs\footnote{These PDFs may not accurately represent the realistic distributions of MWDs, since there are only few hundreds of MWD samples available . Nevertheless, we could treat them as an approximate estimate with a limited database.}.

\begin{figure}	
	\centering
	$$\includegraphics[width=0.45\textwidth]{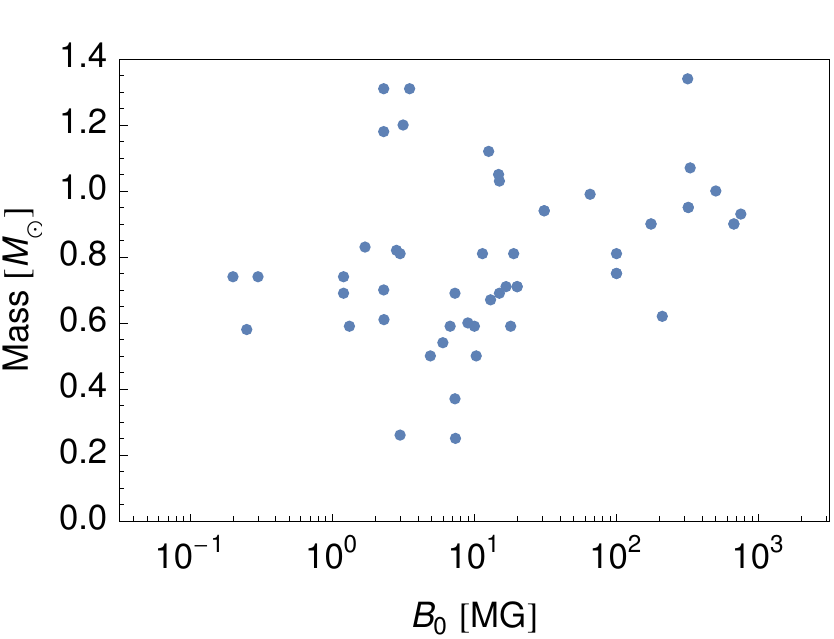}~~~~~
	\includegraphics[width=0.47\textwidth]{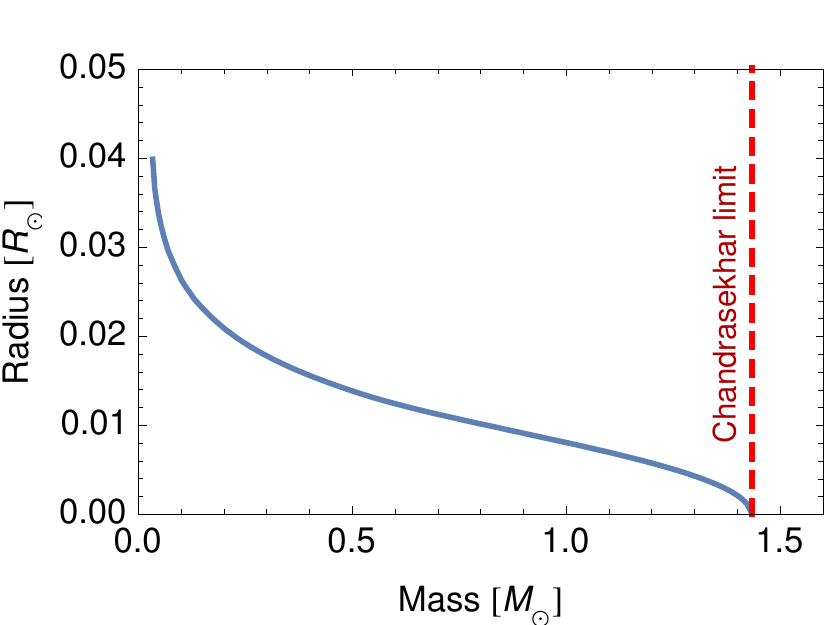}
	$$
	\vspace{-1cm}
	\caption{\em
	On the left: the scatter plot of magnetic field strength $B_0$ and mass of MWDs $M_\text{WD}$ \cite{2015SSRv,Brinkworth_2013}. On the right: the Mass-Radius relationship for a
	WD (solid blue line), while the red dashed line is the Chandrasekhar limit $\sim 1.44~M_\odot$.}
	\label{figs2}
\end{figure}

Before using these PDFs, it is important to examine the correlation between these two variables. In Fig. \ref{figs2} (left), we show the scatter plot of magnetic field strength and mass of MWDs (these data are taken from Table. 1 of Ref. \cite{2015SSRv,Brinkworth_2013}). Based on these data, we can calculate a Pearson correlation coefficient of $\sim$ 0.29, thus we may approximately treat the $R_\text{WD}$ and $B_0$ as independent variables.
Once the MWD mass is given, its radius is determined by the equation of state of MWDs as shown in the right panel of Fig. \ref{figs2} \cite{1935MNRAS,2012.01242}, where the red dashed line represents the Chandrasekhar limit $\sim 1.44~M_\odot$.

\subsection{The properties of neutron stars}
\label{sec:NSproperty}

Similar to MWDs, in order to calculate $r_{c}^\text{NS}$ and $S_{a\gamma}^\text{NS}$, there are also five parameters associated with each NS needed to be determined (see Eq. \eqref{resrNS} and \eqref{eq:NSsingal}).
However, unlike in the case of MWDs, there are still many uncertainties in the equation of state of NSs. Specifically, for a typical solar mass NS, different equations of state give a radius of 10 to 15 km \cite{Lattimer_2001}. For simplicity, we fix the mass and radius of all NS to be the average value, i.e. $M_\text{NS} = 1.44 M_\odot$ \cite{1309.6635}, $R_\text{Ns} = 10.3$ km \cite{1505.05155}. Note that a small deviations from these values do not significantly affect the final results.

As for the other three NS parameters: magnetic field $B_0$, rotation period $P$, and misalignment angle $\alpha$, they are determined by the NS evolution model.
Following Ref. \cite{FaucherGiguere:2005ny,2010MNRAS.401.2675P,1811.01020}, we construct two different NS evolution models, which are labeled as model 1 and model 2. The key difference is that the model 2 includes magnetic field decay, while model 1 does not.
\begin{table}[!t]
	\centering
	\begin{tabular}{c|cccc}
		\hline\hline
		& $~\langle \log (B_\text{in}/G) \rangle~$ & $~\sigma_{\log (B_\text{in}/G)}~$ & $~\langle P_\text{in} \rangle$ [s]~ & $~\sigma_{{P_\text{in}}}$ [s] ~ \\
	
		\hline
		~~Model 1~~ & 12.95 & 0.55 & 0.3 & 0.15 \\
		~~Model 2~~ & 13.25 & 0.6 & 0.25 & 0.1\\
		\hline\hline
	\end{tabular}
\caption{Initial distribution parameters for the magnetic field strength and periods of NS models 1 and model 2 \cite{FaucherGiguere:2005ny,2010MNRAS.401.2675P}.}
\label{tab: IC}
\end{table}

For these two models, the initial value of logarithm of the magnetic field $\log B_\text{in}$ and rotation period $P_\text{in}$ follow a normal distribution. The best-fit central value as well as the standard deviation of $\log B_\text{in}$ and $P_\text{in}$ for models 1 and model 2 are listed in Table. \ref{tab: IC}.
The initial misalignment angle $\alpha_0$ between the NS rotation and magnetic axis follows the simple geometric distribution $p(\alpha_\text{in}) =\frac{1}{2} \sin(\alpha_\text{in})$ with $\alpha_\text{in} \in [0,\pi]$.
However, we find that as the evolution goes on this alignment angle exponentially approaches zero, and thus turns out to play only a minor role in determining the properties of the final NS population.

For model 1 and model 2, the evolution of $P(t)$ and $\alpha(t)$ has the same form, which are given by \cite{1970ApL,1311.1513,1811.01020}
\begin{equation}
		P(t) P'(t)  =\left\{
		\begin{aligned}
			 &\frac{2}{3} \frac{P_\text{in}^2}{\tau_\text{in}} \qquad \qquad  \qquad  \text{for active NSs} \\
			&\frac{2}{3} \frac{P_\text{in}^2}{\tau_\text{in}} \sin^2 \alpha(t) \qquad\hspace{0.09cm} \text{for dead NSs} \\
		\end{aligned},
		\right.
		\label{eq:pt}
\end{equation}
\begin{equation}
		\frac{d}{dt} \text{~log~} \text{sin} \alpha(t)  = -\frac{2}{3} \frac{\text{cos}^2 \alpha_\text{in}}{\tau_\text{in}},
		\label{eq:alpha}
\end{equation}
where
\begin{equation}
	\tau_\text{in} = {I \over \pi \mu^2 f_0^2} \approx 8904 \left( { 10^{12} \, \, \text{G} \over B_0 } \right)^2 \left( {P_\text{in} \over 0.01} \right)^2 \, \, \text{yr}
	\label{eq:tau0}
\end{equation}
is the "decaying" time scale of the misalignment angle $\alpha(t)$.
From Eq. \eqref{eq:alpha}, we find that sin$\alpha(t)$ approaches zero exponentially. This means that the magnetic axis tends to be either parallel or antiparallel to rotation axis, depending on whether the initial misalignment angle $\alpha_\text{in}$ is less or greater than $\pi/2$.
The evolution of $P(t)$ depends on the state of the NS (see Eq. \eqref{eq:pt}). Again, here we have adopted the approximate criteria that the active NSs fulfill $B_0/P^2 > 0.34\times10^{12} \text{~G s}^{-2}$ \cite{Bhattacharya1992}.
In particular, once a NS becomes inactive and $\alpha(t) \to 0$, the evolution of $P(t)$ stops, making it unlikely that the final rotation period will be much larger than the initial value $P_\text{in}$.
Besides, for model 1 the $B_0$ in Eq. \eqref{eq:tau0} is a constant (is equal to $B_\text{in}$), which means  $P(t)$ as well as $\alpha(t)$ in this case can be solved analytically \cite{1811.01020}.

In model 2, the $B_0$ in Eq. \eqref{eq:tau0} is no longer a constant but also evolves over time, and more specifically, decays as it evolves. In general, there are three main decaying channels: Ohmic dissipation, ambipolar diffusion, and Hall drift \cite{1992ApJ,1811.01020}. If a NS is born with a high magnetic field $B_0\gtrsim 10^{16} ~\text{G}$ and high core temperature $T_\text{core} \gtrsim 10^9 ~\text{K}$, the dominant dissipate is through ambipolar diffusion \cite{1992ApJ}, while at low temperature the dissipate process is through a combination of Hall drift and Ohmic heating \cite{astro-ph/0510396}. The evolution of magnetic field $B_0(t)$ is given by \cite{1811.01020}
\begin{equation}
	{d B_0 \over dt}  =-B_0 \left[ {1 \over \tau_\text{ohm}} + \left({B_0 \over B_\text{in}} \right)^2 {1 \over \tau_\text{ambip}} \right],
	\label{eq:Bevo}
\end{equation}
where
\begin{equation}
	\tau_\mathrm{ohm} \sim \frac{1.8\times10^9}{Q_\mathrm{imp}}\mathrm{yr},~~~~ \tau_\mathrm{ambip} \sim 3\times 10^9\left(\frac{t_\text{E}}{10^6 ~\text{yr}}\right)^{-1/3}\left(\frac{B_\text{in}}{10^{12}~\text{G}}\right)^{-2}\mathrm{yr} \,,
\end{equation}
represent the timescale for Hall drift and Ohmic heating, and ambipolar diffusion, respectively. $Q_\mathrm{imp}$ is the impurity in NS's crust, $t_\text{E}$ is the age of the individual NS. As in Ref. \cite{1811.01020}, we randomly assign a value $Q_\mathrm{imp}$ for a log-flat distribution over a range $\left[10^{-3},~10\right]$. For $t_\text{E}$, since we have adopted the assumption that NS-formation rate is a constant over the whole life of $\omega$ Cen ($\sim$ 11.52 Gyr), we just assign a random number within $0\sim11.52~\text{Gyr}$ for each individual NS.

\begin{figure}
	\centering
	$$\includegraphics[width=0.47\textwidth]{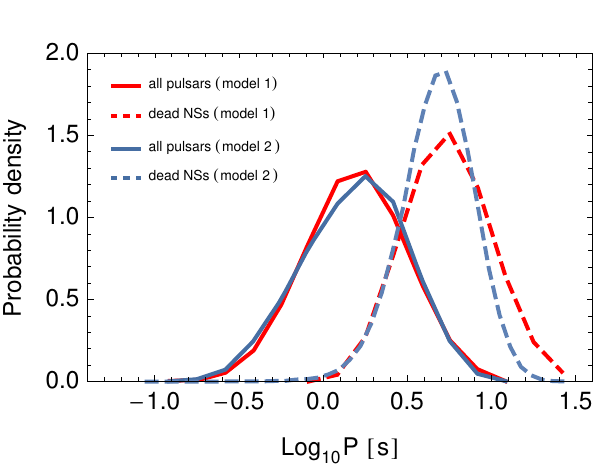}~~~~
	\includegraphics[width=0.47\textwidth]{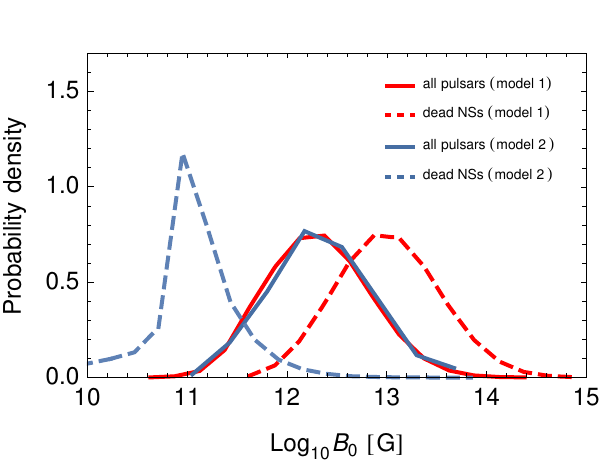}
	$$
	\vspace{-1cm}
	\caption{\em
		On the left: the normalized distribution of rotation period of the evolved NSs. The red and blue lines represent the results of model1 and model 2, respectively. The dashed lines represent the dead NSs, while the active NSs are indicated by solid lines. On the right: as in the left panel, but for the magnetic filed strength.}
	\label{figs3}
\end{figure}

By solving  Eq. \eqref{eq:pt} $\sim$ \eqref{eq:alpha}, and \eqref{eq:Bevo}, we can get the distribution of the parameters of evolved NSs.
In Fig. \ref{figs3} we demonstrate the distribution of rotation period (left) as well as the magnetic field strength (right) for both model 1 (red lines) and model 2 (blue lines).
The solid and dashed lines represent the results of active and dead NSs, respectively.
We find that the distribution of $P$ and $B_0$ of active NSs agree well between the model 1 and model 2. This is reasonable, since the NS models are turned to match the properties of the active pulsars.
Besides, due to the attenuation of the magnetic field in model 2, the magnetic field of the dead NS in model 2 is almost two orders of magnitude smaller than that in model 1.
According to our simulations, the ratio of dead NS is predominant ($\sim 99.7\%$) in both models, so such a huge gap in magnetic fields would cause model 1 to produce a stronger radio signal than model 2 (see Sec. \ref{sec:res}).

\section{The dark matter and compact star distributions in $\omega$ Cen}
\label{sec:starDM}
In addition to the parameters of CSs, it is also crucial to know the properties of $\omega$ Cen, such as the spatial and velocity dispersion distribution of the CSs, the DM profile and so on. Combining all these parameters, we can compute the integrated signal from axion-photon conversion over the ensemble of CSs in $\omega$ Cen. Moreover, we also discuss the spectral characteristics of $\omega$ Cen, which is important for the signal detection on SKA1 and LOFAR (see Sce. \ref{sec:telescope}).

\subsection{The dark matter in $\omega$ Cen}
\label{sec:31}
As we mentioned in Sec. \ref{sec:intro}, many astrophysical observations show that the $\omega$ Cen is unique and may be a remnant core of a dwarf galaxy, so the DM density in $\omega$ Cen might be high. In addition, Ref. \cite{1907.08564} and \cite{Reynoso_Cordova_2021} show that  DM annihilation scenario can indeed explain the  Fermi-LAT data on $\gamma$-ray emission from the direction of $\omega$ Cen \cite{Abdo_2010}. As shown in Ref. \cite{1907.08564}, the mass distribution in $\omega$ Cen can be obtained by solving spherical Jeans equation with assuming that the gravitational potential contains three components of mass: luminous stars, a central black hole, and non-luminous DM. By fitting the tangential velocity dispersion data of stars \cite{Watkins_2015}, the hypothesis that $\omega$ Cen contains no DM is ruled out with $p<0.05$ \cite{1907.08564}.
To model the DM distribution in $\omega$ Cen, we adopt the the Navarro-Frenk-White (NFW) profile \cite{Navarro:1995iw}, which is given by
\begin{equation}
	\rho_\text{NFW} = \rho_s\left(\frac{r}{r_s}\right)^{-1}\left(1+\frac{r}{r_s}\right)^{-2},
	\label{eq:NFW}
\end{equation}
where $\rho_s$ and $r_s$ are the scale density and radius, respectively. For these two parameters, we take benchmark values as $r_s = 1.63 $ pc and $\rho_s = 7650.59 ~M_\odot \text{pc}^{-3}$\footnote{The value of $r_s$ is the median value in Ref. \cite{1907.08564}, while the value of $\rho_s$ is derived by demanding the $J$-factor of $\omega$ Cen is also the corresponding median value $J = 10^{22.1}~ \text{GeV}^2 \text{cm}^{-5}$.}. With these parameters we can estimate the total DM mass in $\omega$ Cen is $\sim 10^6~ M_\odot$.

In order to investigate the impact of NFW profile on the final result, here we also adopt another two sets of parameters: $r_s = 1.0 $ pc, $\rho_s = 27860.5 ~M_\odot \text{pc}^{-3}$, and $r_s = 2.0 $ pc, $\rho_s = 4391.65 ~M_\odot \text{pc}^{-3}$.
With these parameters the total DM mass in $\omega$ Cen is fixed to be $10^6~ M_\odot$ so that the stellar mass of $\omega$ Cen is  $\sim 3 \times 10^6 ~M_\odot$. This is consistent with the scaled result based on the $N$-body simulation of GCs \cite{Kremer_2020}, as mentioned in Sec. III B.
Besides, all these parameters as well as the corresponding $J$-factor and DM mass inside optical half-light radius of $\omega$ Cen are within one standard deviation of the best fit as shown in \cite{1907.08564}.
We find that for different NFW parameters, their impact on the limit of $g_{a\gamma}$ (see Fig. \ref{figs6}) is less than 10\% compared to the benchmark values.
As for other different DM profile, e.g. the Burkert \cite{Burkert_1995} and the Moore \cite{Moore_1999}, a detailed study on the GC kinematics is needed. This is beyond the scope of this work.
Therefore, in the following sections we only show the results of benchmark NFW profile.

\subsection{The population and spatial distribution of compact stars in  $\omega$ Cen}
\label{sec:32}

The simulation of the formation of WDs and NSs in GCs has been studied in Ref. \cite{Kremer_2020,arXiv210602643,2104.11751,2103.05033}. Ref. \cite{Kremer_2020} demonstrates a set 148 independent $N$-body simulations of GCs, which is derived by using the Cluster Monte Carlo code. These results roughly cover the full range of cluster properties in  the Milky Way.
For each simulation, there are four initial cluster parameters: virial radius $r_v$ that indicates the initial cluster size, total number of particles (single stars plus binaries) $N$, galactocentric distance $R_\text{gc}$, and metallicity $Z$. Note that for $N\gtrsim 10^6$, the direct $N$-body integration becomes extremely computationally expensive \cite{stw274}. As a practical approach, we choose the model that is most similar to $\omega$ Cen and then scale the corresponding simulation results to the realistic $\omega$ Cen sizes \cite{Baumgardt}.

The age of $\omega$ Cen is 11.52 Gyr and its metallicity is $[\text{Fe}/\text{H}]\approx-1.35$ (or $Z_{\omega} = 0.045~ Z_\odot$) \cite{1211.4399,omegaFeH}. The right ascension and declination of $\omega$ Cen are	13h 26m 47.28s and $-47^{\circ}$ 28' 46.1"\cite{2010AJ}, so its galactocentric distance $R_\text{gc}^\omega$ is $\sim$ 6.5 kpc. With these properties, we find that the model 52 (with $N = 16\times10^6$, $r_v = 1$ pc, $R_\text{gc} = 8$ kpc, and $Z = 0.01~Z_\odot$) in Ref. \cite{Kremer_2020} is the best option, which is consistent with the choice in Ref. \cite{2103.05033}.

There are about $\sim 10^7$ stars in $\omega$ Cen. With the given star number and stellar mass in the simulation, we can obtain the stellar mass in $\omega$ Cen $\sim 2.94\times10^6~M_\odot$ via scaling. This is consistent with the value $\sim2.92\times 10^6 ~M_\odot$, which is obtained by subtracting the DM mass given by the benchmark NFW profile from the total mass.
After the scaling, we can also conclude that in $\omega$ Cen there are about 1084100 WDs (means the number of MWDs with $B_0\gtrsim0.1~\text{MG}$ is $\sim$ 102990) and 12531 NSs. Note that for the bright low-mass X-ray binaries (LMXBs) as well as cataclysmic variables (CVs), the accretion is very active and may change the plasma distribution around the CSs and make the axion-photon conversion calculation much more complicated. Fortunately, according to the simulation in Ref. \cite{Kremer_2020}, both LMXBs and CVs represent only a very small fraction ($<1\%$) of the total number of CSs, so these special cases can be safely ignored.

\begin{figure}
	
	\centering
	$$\includegraphics[width=0.48\textwidth]{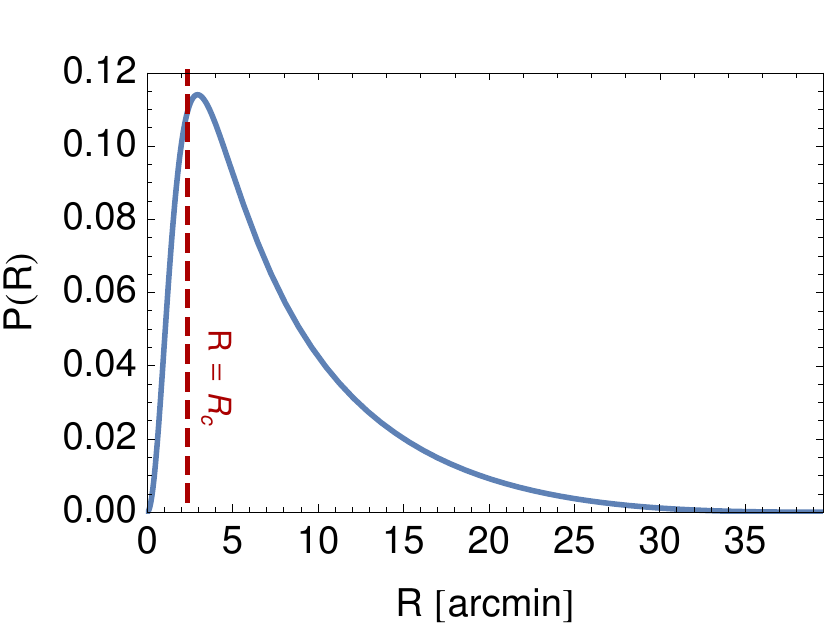}~~~~
	\includegraphics[width=0.47\textwidth]{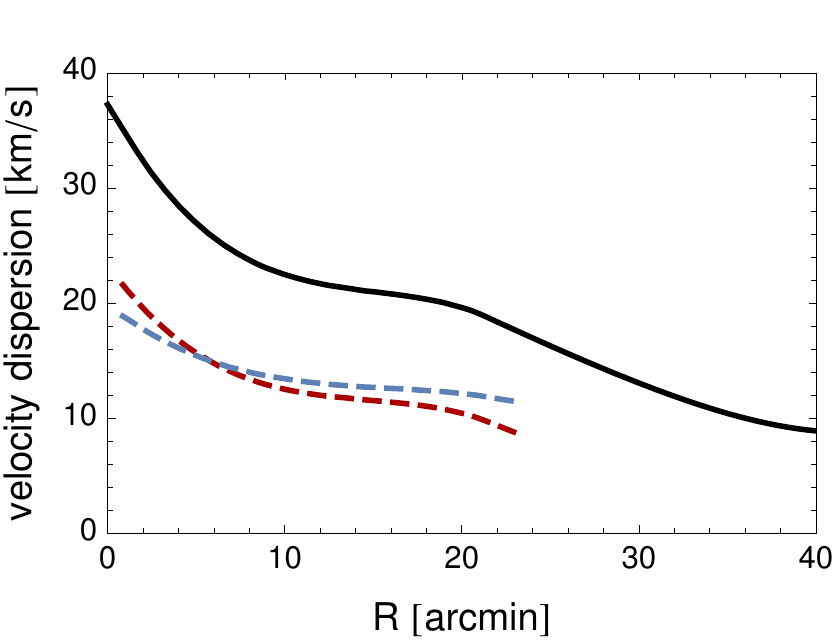}
	$$
	\vspace{-1cm}
	\caption{\em
	On the left: the spatial probability distribution of stars in $\omega$ Cen for a King profile with $R_c = 141.676''$, $c=1.224$ (solid blue line), while the red dashed line indicates the position of $R_c$ \cite{1211.4399}. On the right: radial (red dashed) and tangential (blue dashed) proper motion dispersion profiles as a function of projected radius \cite{1211.4399}. The black solid line is the total velocity dispersion $\langle v^2 \rangle$. Note that for $R>23''$ region, we use the linear interpolation.}
	\label{figs4}
\end{figure}

Given that the DM density is very location-dependent (see Eq. \eqref{eq:NFW}),  the spatial distribution of CSs in $\omega$ Cen is crucial for the final result.
It is well known that the distribution of stars in GCs can be well fitted by the King model, which is derived by solving the collisionless Liouville equation for a given
velocity distribution \cite{1966AJ}. The King model profile is described by two parameters: the core radius $R_c$ and the concentration parameter $c = \text{log}(R_t/R_c)$, where $R_t$ is tidal radius. These two parameters can be derived by fitting the normalized surface number density profile of the given GC.  For $\omega$ Cen, these two parameters are: $R_c = 141.676''$, $c = 1.224$ \cite{1211.4399}.
After determining the King model, it is straightforward to derive the spatial distribution of stars in $\omega$ Cen. As shown in Fig. \ref{figs4} (left), the solid blue line represents the desired spatial PDF, while the red dashed line represents the location of $R_c$. In our analysis, we adopt the assumption that all the CSs fulfill this spatial distribution.

%

\subsection{Radio spectrum of CS ensemble in $\omega$ Cen}
\label{sec:33}

In comparison with the single CS case, the axion-induced radio spectrum of GC is more complicated due to the velocity dispersion of CSs \cite{1811.01020}.
In Fig. \ref{figs4} (right) we show the radial ($\sqrt{\langle v_R^2\rangle}$, red dashed) and tangential ($\sqrt{\langle v_T^2\rangle}$, blue dashed) proper motion dispersion as a function of projected
radius of $\omega$ Cen. We find that both profiles are comparable and varies from $\sim10$ km/s at distance $\sim 23''$ to $\sim20$ km/s at center region \cite{1211.4399}. The black solid line represents the total velocity dispersion $\langle v^2\rangle$, which is derived from $\langle v^2\rangle= \frac{3}{2}\langle v_R^2 + v_T^2 \rangle$ \cite{1989ApJ}.  The values in $R>23''$ region are derived by using the linear interpolation.
Here we take the simplifying assumption that both DM and CSs have the same homogeneous and isotropic Maxwell-Boltzmann velocity distribution with $v_0^2 \sim \frac{2}{3} \langle v^2\rangle$. In this case, the DM velocity dispersion is clearly position dependent.

For an isolated CS, the center frequency of the spectrum in CS rest frame is around $f_c \sim m_a/(2\pi)$ with a bandwidth is $B_\text{CS} \sim v_0^2 m_a/(2\pi)$, which originates from the velocity dispersion of DM. However, for an ensemble of CSs, each CS has different $f_c$ in observer's frame due the Doppler shift. In observer's frame, the observed center frequency of each CS  $\tilde{f_c}$ is given by $\tilde{f_c}=f_c\sqrt{\frac{1-v_\text{LOS}}{1+v_\text{LOS}}}\approx f_c (1-v_\text{LOS})$. $v_\text{LOS}$ is the projection of CS's velocity along the line of sight and is defined to be positive for the CS moving away from the observer.
According to the characteristics of Maxwell-Boltzmann velocity distribution, we find the $P(|v_\text{LOS}|<3 v_0)\approx1$. Therefore, in order to include all CS's contribution, we set the GC bandwidth as $B_\text{GC}\sim 6 v_0^\text{max} f_c \sim 6\times10^{-4} m_a/(2\pi)$.

\section{Radio telescope sensitivity}
\label{sec:telescope}
In this section, we introduce several key parameters of SKA1 and LOFAR and describe how to estimate their detection sensitivity. The SKA1 consists of a low-frequency
aperture array (SKA1 Low) and a middle frequency aperture array (SKA1 Mid).
The LOFAR contains a low-band antennas (LBAs) and a
high-band antennas (HBAs). In Talbe. \ref{tab:AeffTsys}, we list the detailed parameters of frequency range, resolution $\mathcal{B}_\text{res}$, and field of view (FoV) for each band. It is not hard to see that SKA1 covers a much wider frequency range $0.35\sim15.3$ GHz, while the LOFAR can detect a lower frequency $\sim 0.03$ GHz. Therefore, the combination of SKA1 and LOFAR can make a more effective detection of axion DM.
\begin{table}[!t]
	\centering
	\begin{tabular}{c|c|ccc}
		\multicolumn{5}{c}{Technical parameters of the SKA1 and LOFAR \cite{skainfo,1305.3550}} \\
		\hline\hline
		\multicolumn{2}{c|}{~Channel~}    & ~~Range [GHz]~ & ~Resolution [kHz]~ & ~FoV [arcmin]~  \\
		\hline
		\multicolumn{2}{c|}{SKA1 LOW}     & 0.05$\sim$0.35 & 1.0              & 327           \\
		\hline
		\multirow{6}{*}{SKA1 MID} & B1    & 0.35$\sim$1.05 & 3.9              & 109           \\
		& B2    & 0.95$\sim$1.76 & 3.9              & 60            \\
		& ~B3$^*$    & 1.65$\sim$3.05 & 9.7              & 42              \\
		& ~B4$^*$    & 2.80$\sim$5.18 & 9.7              & 42              \\
		& B5a   & 4.6$\sim$8.5   & 9.7              & 12.5          \\
		& B5b   & 8.3$\sim$15.3  & 9.7              & 6.7           \\
		\hline
		\multicolumn{2}{c|}{LOFAR LBA}   & 0.03$\sim$0.08 & 195              & 470.9           \\
		\multicolumn{2}{c|}{LOFAR HBA}   & 0.11$\sim$0.24 & 195              & 94.8           \\
		\hline\hline
	\end{tabular}
	\caption{The frequency range, spectral resolution, and the FoV
		of the different frequency band for SKA1 and LOFAR. Note that B3 and B4 band are not formally part of the design baseline, so the FoV of these two bands are set to be the mean value of  SKA1 MID \cite{skadesign}. }
	\label{tab:AeffTsys}
\end{table}

\begin{figure}
	\centering
	$$\includegraphics[width=0.47\textwidth]{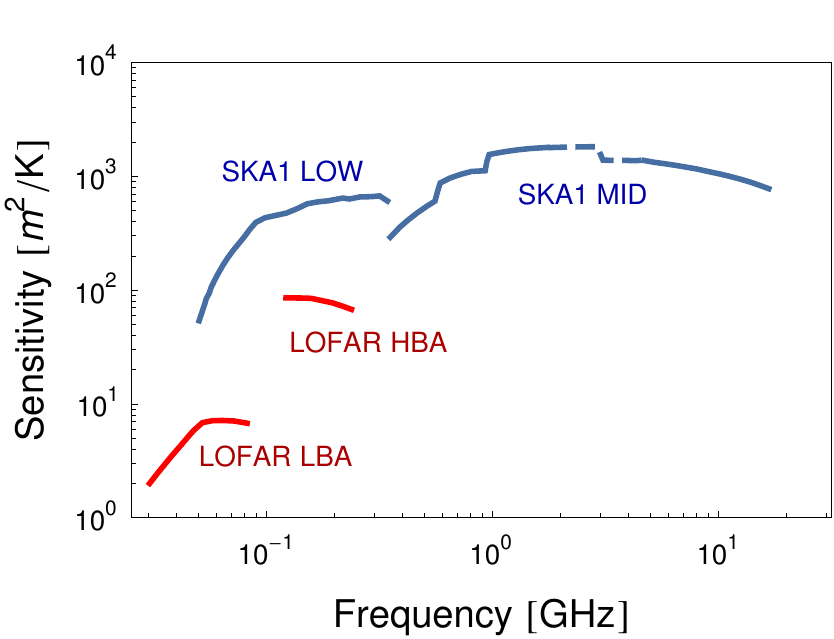}~~~~~
	\includegraphics[width=0.46\textwidth]{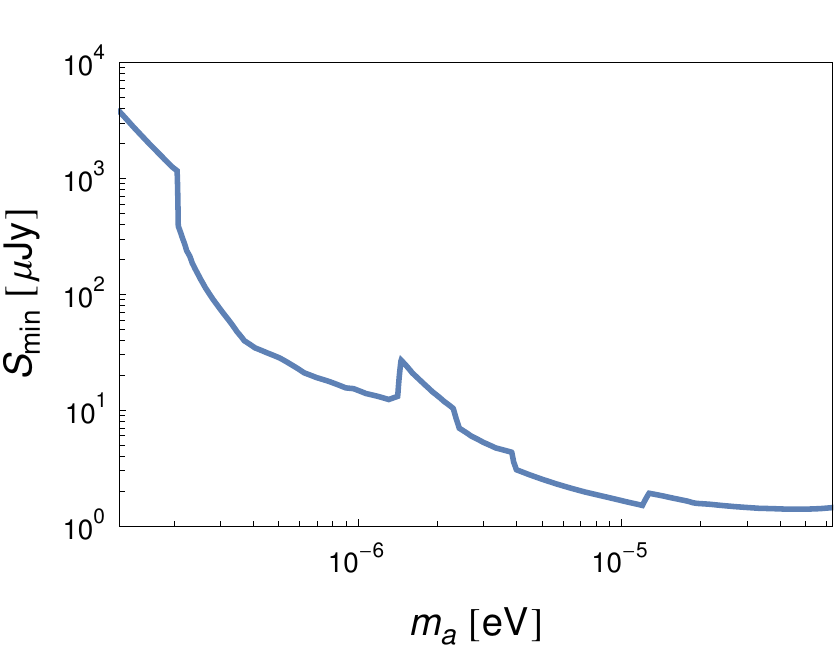}
	$$
	\vspace{-1cm}
	\caption{\em \label{fig:sensiti}
		On the left: the sensitivity of SKA1 (blue) and LOFAR (red) at different frequency bands \cite{skapro}. Note that the dashed blue line represents the B3 and B4 bands. On the right: the combined $S_\text{min}$ of SKA1 and LOFAR. 
	}
	\label{figs5}
\end{figure}

For the detection purpose, a useful physical quantity is the minimal detectable flux density, which is given by \cite{1804.03145,2101.02585}:
\begin{equation}
	S_{\min} = \frac{\rm SEFD }{\eta_s \sqrt{n_{\rm pol}  \, {\cal B} ~t_{\rm obs}} }\ ,
	\label{eq:Smin1}
\end{equation}
where
\begin{equation}
	{\rm SEFD } = \frac{2 k_B }{ A_{\rm eff}/T_{\rm sys} }
	\label{eq:SEFD}
\end{equation}
is the system-equivalent flux density, $\eta_s$ is the system efficiency, $n_{\rm pol} =2$ is the number of polarization, $\mathcal{B} = \text{max}\{B_\text{GC}, B_\text{res}\}$ is the optimized bandwidth for the GC,
$t_{\rm obs}$ is the observation time,
$k_B$ is the Boltzmann constant, $T_{\rm sys}$ is the antenna system temperature, and
$A_{\rm eff}$ is the antenna effective area of the array.
In Fig. \ref{figs5} (left), we show the sensitivity ($A_{\rm eff}/T_{\rm sys}$) of SKA1 and LOFAR at different frequency bands \cite{skapro}.

In this analysis, we take $\eta_s = 0.9$ for SKA1~\cite{skainfo} and $\eta_s = 1.0$ for LOFAR~\cite{1305.3550}. The $t_\text{obs}$ is set to be 100 hours as a benchmark. With these parameters, we can calculate the combined $S_\text{min}$ of SKA1 and LOFAR for different frequency band (see Fig. \ref{figs5} (right)).
Note that the telescope's FoV gradually decreases as the frequency increases,  so for each frequency band, we need to carefully subtract the contribution of CSs outside the corresponding FoV.

\section{Results}
\label{sec:res}

For the purpose of calculating the total energy flux density of all CSs in $\omega$ Cen,
an important step is to calculate the energy flux density of an individual CS. The detailed derivation of this part of the calculation can be found in the previous literatures \cite{1804.03145,1903.05088,2101.02585}.
Following Ref. \cite{1804.03145,2101.02585}, the axion-induced radio flux density from a single MWD and NS are given by:
\begin{align}
	\label{eq:WDsingal}
	S_{a\gamma}^\text{WD} & \simeq 0.003 ~\mu\text{Jy}\left(\frac{\xi(r_c)^2}{\sin^2(\tilde{\theta})}\frac{3 \cos^2(\theta)+1}{|\xi'(r_c)H_\text{cor}-3\xi(r_c)|}\right)
	\left(\frac{r_c}{10^4 ~\text{km}}\right)^{-\frac{9}{2}}
	\left(\frac{H_\text{cor}}{10^2 ~\text{km}}\right)
	\left(\frac{M_\text{NS}}{M_\odot}\right)^{\frac{1}{2}} \left(\frac{\rho_\text{DM}}{10^3 ~\text{GeV}/\text{cm}^3}\right)
	\notag\\
	&\times
	\left(\frac{R_\text{NS}}{10^4 ~\text{km}}\right)^{6}
	\left(\frac{10~ \text{km}/\text{s}}{v_0}\right)
	\left(\frac{g_{a\gamma}}{10^{-12}~\text{GeV}^{-1}}\right)^2\left(\frac{B_0}{10^{8}~\text{G}}\right)^{2}\left(\frac{1~\mu\text{eV}}{m_a}\right)\left(\frac{d}{1~\text{kpc}}\right)^{-2}\left(\frac{\mathcal{B}}{1~\text{kHz}}\right)^{-1},
\end{align}

\begin{align}
	\label{eq:NSsingal}
	S_{a\gamma}^\text{NS} & \simeq 2.96 ~\mu\text{Jy}\left(\frac{\xi(r_c)^2}{\sin(\tilde{\theta})^2}\frac{3 \cos(\theta)^2+1}{|\xi'(r_c)r_c-3\xi(r_c)|}\right)
	\left(\frac{r_c}{10^2 ~\text{km}}\right)^{-\frac{7}{2}}
	\left(\frac{M_\text{NS}}{M_\odot}\right)^{\frac{1}{2}} \left(\frac{\rho_\text{DM}}{10^3 ~\text{GeV}/\text{cm}^3}\right)
	\left(\frac{R_\text{NS}}{10 ~\text{km}}\right)^{6}
	\notag\\
	&\times
	\left(\frac{10~ \text{km}/\text{s}}{v_0}\right)
	\left(\frac{g_{a\gamma}}{10^{-12}~\text{GeV}^{-1}}\right)^2\left(\frac{B_0}{10^{14}~\text{G}}\right)^{2}\left(\frac{1~\mu\text{eV}}{m_a}\right)\left(\frac{d}{1~\text{kpc}}\right)^{-2}\left(\frac{\mathcal{B}}{1~\text{kHz}}\right)^{-1},
\end{align}
where
\begin{equation}
	\xi(r) = {\sin^2 \tilde \theta \over 1 - {\omega_p^2(r) \over \omega^2(r)} \cos^2 \tilde \theta }, ~~~\text{with } \omega_p^2(r) = \left\{
	\begin{aligned}
		& m_a^2 ~\text{exp}\left(-\frac{r-r_c}{H_\text{cor}}\right)  \hspace{0.4cm} \text{for MWDs} \\
		&m_a^2 \left(\frac{r_c}{r}\right)^3 \hspace{2.02cm} \text{for NSs} \\
	\end{aligned},
    \right.
\end{equation}
\begin{equation}
	H_{\text{cor}}=\frac{2 k_\text{B} T_\text{cor}}{m_\text{p} g}=21.90 \left(\frac{T_\text{cor}}{10^6 ~\text{K}}\right) \left(\frac{M_\text{WD}}{M_\odot}\right)\left(\frac{R_\text{WD}}{10^4 ~\text{km}}\right)^{-2} \text{km}
	\label{eq:hcor}
\end{equation}
is the scale height of the isothermal corona of MWDs \cite{2101.02585}, and $d$ represents the distance from the CS to us.

Combining Eq. \eqref{resrWD}$\sim$\eqref{resrNS} and Eq. \eqref{eq:WDsingal}$\sim$\eqref{eq:hcor}, we can calculate the specific energy flux density for any NS or MWD sample in our simulation.
Therefore, the total energy flux density of all CSs in the $\omega$ Cen can be expressed as:
\begin{equation}
	S_{a\gamma}^\text{total} =\left( \sum_{N_\text{NS}} S_{a\gamma}^\text{NS}  + \sum_{N_\text{WD}} S_{a\gamma}^\text{WD}\right)\times \Theta\left(\frac{\text{FoV}}{2}-P_\text{CS}\right),
	\label{eq:smintotal}
\end{equation}
where the $N_\text{NS}$ and $N_\text{WD}$ represent the number of NSs and MWDs, $P_\text{CS}$ is the projected position of each CS on the plane of sky. The step function ensures that in each frequency band, only the contribution of the CS in the corresponding FoV will be counted.
Another thing to note is that the second factor in Eq. \eqref{eq:WDsingal} and \eqref{eq:NSsingal} is dimensionless and roughly an order one number.
Therefore, some rough qualitative estimates can be obtained by ignoring this item. Roughly speaking, we have (see the Eq. (20)$\sim$(21) in Ref. \cite{2101.02585}):
\begin{equation}
    S_{a\gamma}^\text{WD} \propto \rho_\text{DM} v_0^{-1} g_{a\gamma}^2 B_0^2 m_a^{-1}d^{-2}, \quad  S_{a\gamma}^\text{NS} \propto \rho_\text{DM} v_0^{-1} g_{a\gamma}^2 P^{7/6} B_0^{5/6} m_a^{4/3}d^{-2}.
    \label{eq:appsim}
\end{equation}
Note that Eq. \eqref{eq:appsim} is very helpful to understand the following results. In summary, the whole calculation process can be reduced to the following steps:
\begin{enumerate}[label=(\roman*)]
	\item generate 102990 MWD and 12531 NS samples. For MWDs, the mass and magnetic field strength are randomly sampled through the distribution functions shown in the Fig. \ref{figs1}, while the radius of each sample is determined by the equation of state (see Fig. \ref{figs2} (right)). For NSs, the final rotation period of each NS is determined by Eq. \eqref{eq:pt}. The magnetic field in model 2 evolves according to Eq. \eqref{eq:Bevo}, while in model 1 is equal to the corresponding initial value $B_\text{in}$;
	\item for each CS, its position in $\omega$ Cen is randomly produced according to the King model (see Fig. \ref{figs4} (left)). The corresponding DM density as well as the velocity dispersion are derived through Eq. \eqref{eq:NFW} and Fig. \ref{figs4} (right), respectively. Besides, for $\theta$, which denotes the angle between the CS's rotation axis and the line-of-sight, is randomly generated based on $P(\theta) = \frac{1}{2} \sin\theta$ with $\theta \in [0,\pi]$;
	\item with the parameters of these CS samples, we can calculate the total energy flux density $S_{a\gamma}^\text{total}$ by using Eq. \eqref{eq:WDsingal}$\sim$\eqref{eq:NSsingal} and \eqref{eq:smintotal};
	\item by comparing $S_{a\gamma}^\text{total}$ with $S_{\min}$, we can finally obtain the limitation of axion-photon coupling coefficient $g_{a\gamma}$ at different axion mass $m_a$.
\end{enumerate}
\begin{figure}
	\centering
	$$\includegraphics[width=0.47\textwidth]{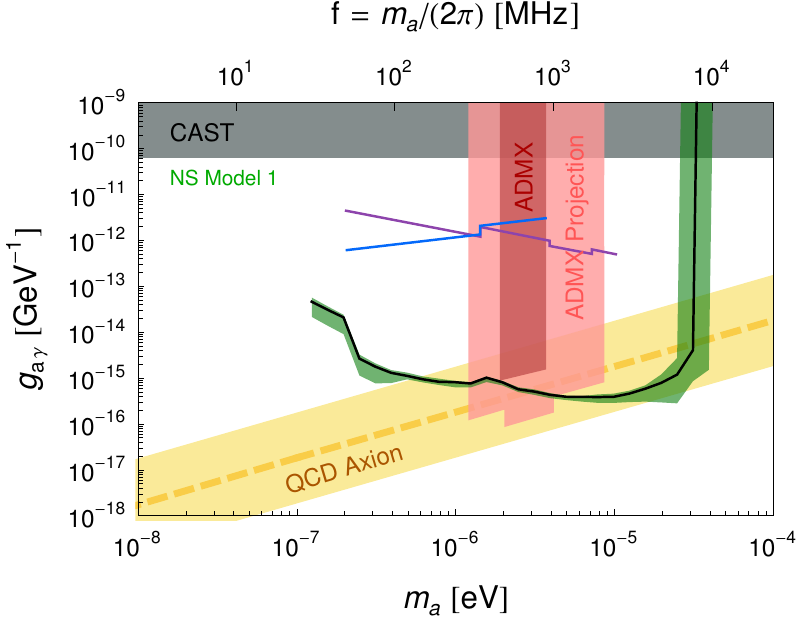}~~~~
	\includegraphics[width=0.47\textwidth]{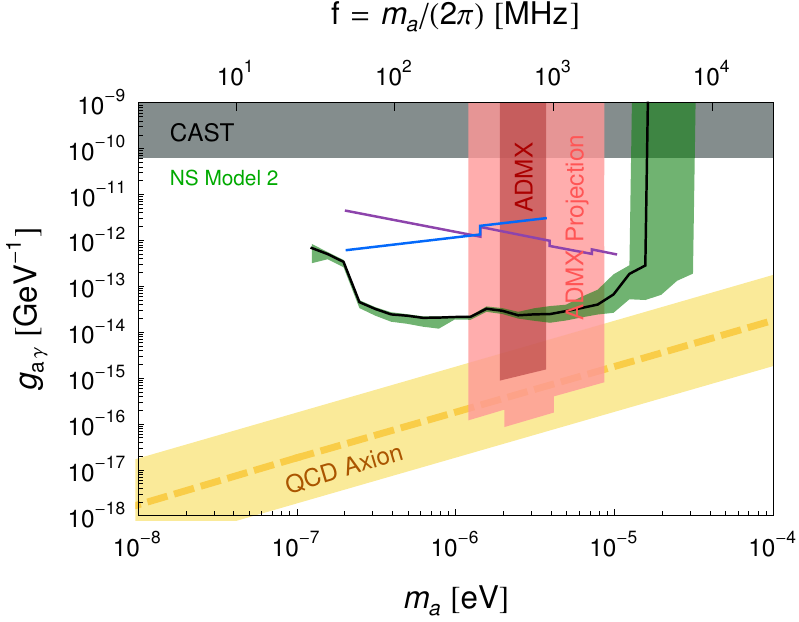}
	$$
	\vspace{-1cm}
	\caption{\em
		The combined projected sensitivity (pure NSs case) to $g_{a\gamma}$ as a function of the axion mass $m_a$ for SKA1 and LOFAR with 100 hours observations of the $\omega$ Cen is shown in the green band. The green band contains ten separate sets of NS samples, and its upper and lower boundaries represent the maximum and minimum values, and the black solid line represent the median value.
		The left panel assumes NS model 1, wile the right panel takes NS model 2. For comparison, the results of the isolated NS RX J0806.4-4123 and MWD WD 2010+310 are shown with purple and blue solid lines. The QCD axion is predicted to lie within the yellow band. The limits set by CAST and ADMX (current and projected) are indicated by the gray and red regions, respectively.}
	\label{figs6}
\end{figure}

In Fig. \ref{figs6} we show the projected sensitivity for the axion-photon coupling $g_{a\gamma}$, taking into account only the contribution of NSs in $\omega$ Cen. The left panel assumes NSs model 1, while the right panel takes NS model 2. The green band show the detection potential of SKA1 and LOFAR with 100 hours of
observation.
To illustrate the effect of statistical fluctuation, we show the results of ten separate NS samples. The upper and lower boundaries of the green band represent the maximum and minimum values of the sensitivity among the ten samples at each $m_a$, respectively. The black solid line represents the median value.
For comparison, we also show the results of isolate MWD WD 2010+310 and NS RX J0806.4-4123 with blue and purple solid lines \cite{2101.02585}. The yellow band represents the typical parameter region predicated by the QCD axion models. The limits set by CAST \cite{1705.02290,1307.1985,1503.00610} and ADMX (current and projected) \cite{PhysRevD.64.092003,PhysRevLett.104.041301,1405.3685} are
indicated by the gray and red regions, respectively .

Compared to the single NS results, we find that for NS model 1 the limitation on $g_{a\gamma}$ has improved by three and a half orders of magnitude, which is mainly due to the enhancement of large DM density in $\omega$ Cen. In addition, the large NS sample number as well as relative small DM velocity dispersion also have a noticeable effect (see Eq. \eqref{eq:appsim}).
According to the calculation results, the top 5\% of NS samples contribute about 90\% of the total flux density of $\omega$ Cen (similar for NS model 2 case).
Typically, these NSs with large contributions have relatively larger magnetic fields and are surrounded by a higher DM density environment, compared to other NSs.
In comparison with model 1, the sensitivity of model 2 is reduced by one and a half orders of magnitude, and this is due to the difference in the magnetic field strength (see Fig. \ref{figs3}). Besides, we also find that the largest detectable axion mass $m_a^\text{max}$ in model 1 is larger than in model 2. For NSs, the $m_a^\text{max}$ can be derived by solving Eq. \eqref{resrNS} with $r_c = R_\text{NS}$, that is
\begin{equation}
	m_a^\text{max} \approx 69.2 \times \big|3 \cos^2 \theta  - 1 \big|^{1/2} \sqrt{\frac{B_0}{10^{14} ~\text{G}}\frac{1 ~\text{s}}{P}} ~\mu\text{eV}.
\end{equation}
Clearly, we can see that this difference is also caused by the difference in NSs' magnetic field (see Fig. \eqref{figs3}). For the $m_a \gtrsim 10~ \mu\text{eV}$ region the green band in model 2 is much wider than in model 1. This is because that the statistical fluctuation of magnetic field with $B_0 \gtrsim 10^{12}$ G in model 2 is much larger.
\begin{figure}
	\centering
	$$\includegraphics[width=0.47\textwidth]{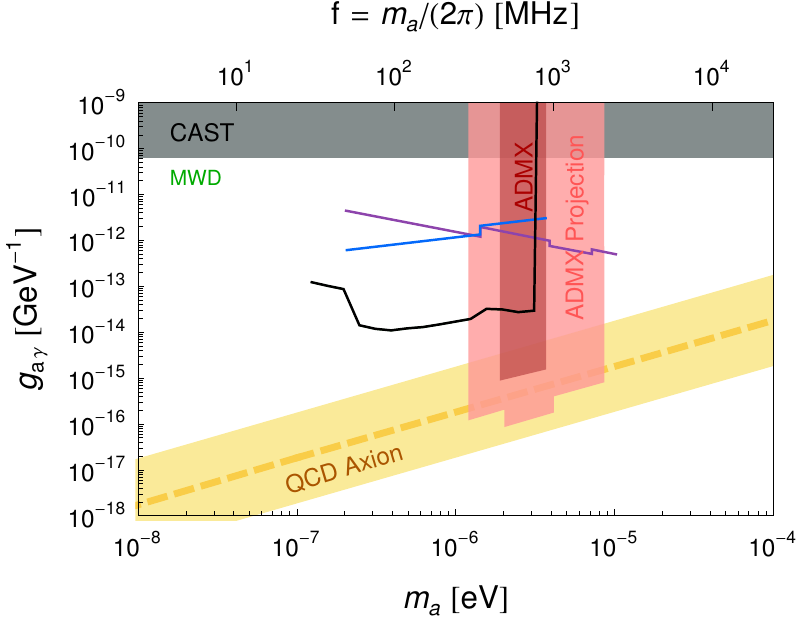}
	$$
	\vspace{-1cm}
	\caption{\em
		As in Fig. \ref{figs6},  except that only the contribution of MWDs is included. Again, ten separate sets of MWD samples are considered. The black solid line represents the median value, while the green band is too narrow to be seen, because the statistical fluctuations are very small.}
	\label{figs7}
\end{figure}

In Fig. \ref{figs7}, we show the projected sensitivity of $\omega$ Cen only considering the contributions of MWDs. Again, ten separate sets of MWD samples are included. Compared to Fig. \ref{figs6}, the green band is extremely narrow and almost invisible. The main reason is that there are so many MWDs, the statistical fluctuation between different sets of MWD samples is small (law of large numbers).
According to Fig, \ref{figs1} (left), there are about $8\%$ MWDs with magnetic fields greater than 100 MG. Therefore, it can be expected that the pure MWDs gives comparable result with pure NS case with model 2. The calculation results show that the top 5\% of MWD samples contribution about 98\% of total flux density.
\begin{figure}
	\centering
	$$\includegraphics[width=0.47\textwidth]{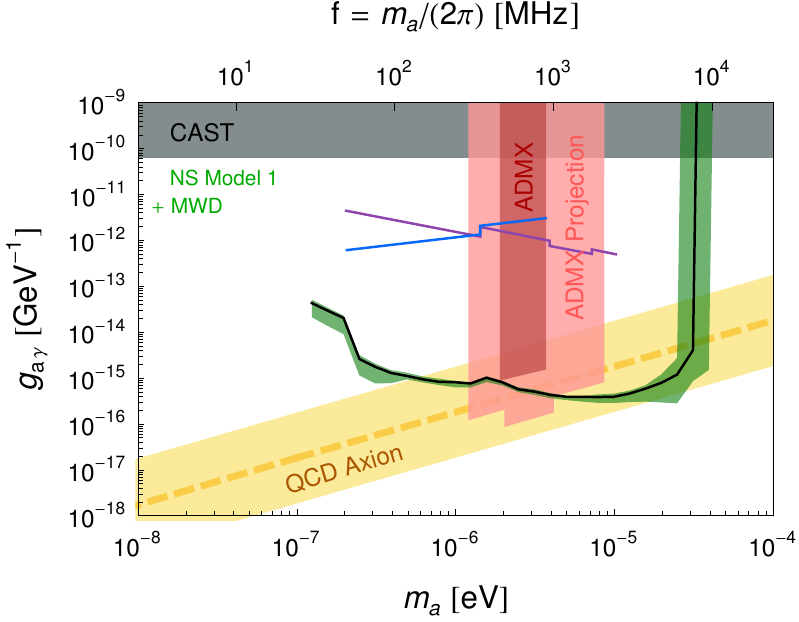}~~~~
	\includegraphics[width=0.47\textwidth]{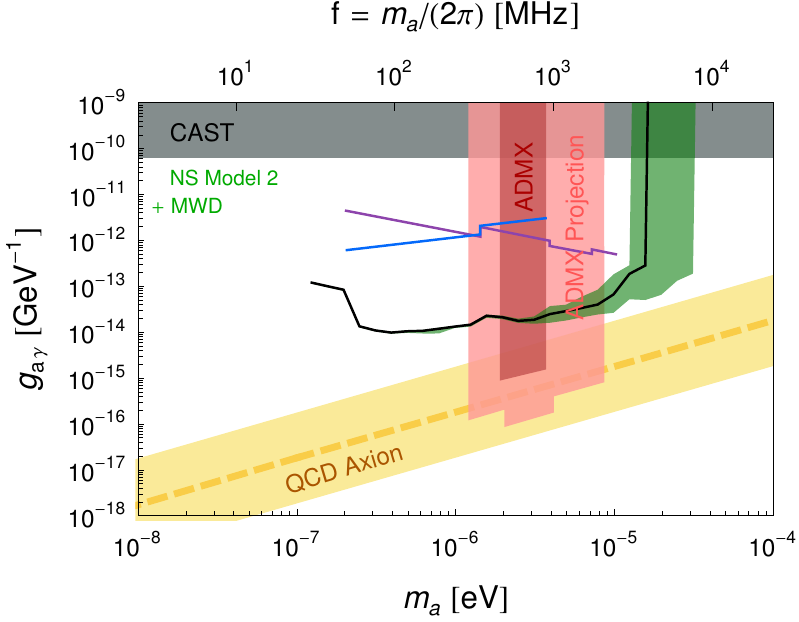}
	$$
	\vspace{-1cm}
	\caption{\em
			As in Fig. \ref{figs6},  except that the contributions of all MWDs and NSs are included. Compared with Fig. \ref{figs8}, the contribution of NS in model 1 is always dominant over the entire mass interval, while for NS model 2, the contribution of MWDs is relatively larger in the region of $m_a\lesssim 2 ~\mu\text{eV}$.}
	\label{figs8}
\end{figure}
Considering that we have chosen the same benchmark corona parameters for all MWDs, so for pure MWDs case, the $m_a^\text{max} \approx 3.7 ~\mu\text{eV}$ is a constant for every individual MWD (see Eq. \eqref{resrWD}).
Even so, for different MWD corona parameters $n_{e0}$ and $T_\text{cor}$, their influence on the final result can also be easily evaluated by using $S_{a\gamma}^\text{WD} \propto n_{e0}^0 T_\text{cor}$ and $m_a^\text{max} \propto n_{e0}^{1/2} T_\text{cor}^0$ \cite{2101.02585}.
In Fig. \ref{figs8}, we show the projected sensitivity of all CSs in $\omega$ Cen.
We find that the contribution of NS in model 1 is always dominant over the entire mass interval, while for NS model 2, the contribution of MWDs is relatively larger in the region of $m_a\lesssim 2 ~\mu\text{eV}$.

\section{Conclusions }
\label{sec:conclusion}
In this work we propose to use the $\omega$ Cen as a probe to detect the axion DM through the radio signals.
As the largest GC in Milky Way, $\omega$ Cen is quite unique and is suggested to be the remnant core of a dwarf galaxy whose outskirts were tidally stripped as it fell into the Milky Way \cite{1907.08564,2019NatAs,Bekki2003}.
Compared to the single NS or MWD, $\omega$ Cen has three compelling advantages: (1) very high DM density; (2) large number of CSs; (3) relatively small DM velocity dispersion.
All of these properties/advantages make the $\omega$ Cen an ideal target for detecting axion DM.

In order to calculate total axion-induced radio signal from $\omega$ Cen, it is necessary to know the distribution of CSs' properties, the number and spatial distribution of CSs, and the DM profile in $\omega$ Cen.
For the properties of MWDs, we get the parameter distribution by using the current available MWDs data with linear interpolation, while for NSs, two evolution models are constructed to derive the parameter of NS ensemble.
The DM distribution in $\omega$ Cen can be derived by solving spherical Jeans equation \cite{1907.08564}.
In our analysis, we adopt the NFW profile with $r_s = 1.63 $ pc, $\rho_s = 7650.59 ~M_\odot \text{pc}^{-3}$. With these parameters we can estimate the total DM mass in $\omega$ Cen is $\sim 10^6~ M_\odot$.

The population of CSs in $\omega$ Cen can be approximately estimated by using the $N$-body simulations of GCs in Ref. \cite{Kremer_2020}. After scaling the evolution results of model 52 with $N = 16\times10^6$, $r_v = 1$ pc, $R_\text{gc} = 8$ kpc, and $Z = 0.01~Z_\odot$ in Ref. \cite{Kremer_2020}, we find that there are about 102990 MWDs with $B_0\gtrsim0.1~\text{MG}$ and 12531 NSs in $\omega$ Cen.
With respect to the spatial distribution of these CSs, we adopt the King model with  $R_c = 141.676''$, $c = 1.224$ \cite{1966AJ,1211.4399}.
Unlike the case of a single CS, the frequency spectra of $\omega$ Cen is much wider due to the Doppler shift effect.
In order to include all CS's contribution, we set the GC bandwidth as $B_\text{GC}\sim 6 v_0^\text{max} f_c \sim 6\times10^{-4} m_a/(2\pi)$.

In Fig. \ref{figs6} $\sim$ Fig. \ref{figs8},  we demonstrate the combined projected sensitivity to $g_{a\gamma}$ on the SKA1 and LOFAR with 100 hours observation. The green band contains ten separate sets of CS samples, and its upper and lower boundaries represent the maximum and minimum values, while the black solid line represent the median value.
In comparison with the single MWD (blue solid) and NS (purple solid) result \cite{1804.03145,2101.02585}, the sensitivity limit of $g_{a\gamma}$ has been improved by two or three and a half orders of magnitude, depending on the choice of NS evolution model.
Note that in some regions, the limitation even reaches the QCD axion scenario.

However, it should be noted that there are still some uncertainties in our calculations, such as the NS evolution model, the DM profile and its velocity dispersion in $\omega$ Cen, number of CSs in $\omega$ Cen, and so on.
In order to better deal with these uncertainties, more observational data (e.g. mass and magnetic filed of MWDs) and more careful numerical simulations (e.g. DM and CSs evolution in $\omega$ Cen) are needed, which are beyond the scope of this paper.
Considering such promising results of $\omega$ Cen, a more detailed study of these uncertainties is deserved and will be left for our future work.

\begin{acknowledgments}
The authors would like to thank Lilia Ferrario and Piero Ullio for helpful discussions.
The work of JWW is supported by the research grant "the Dark Universe: A Synergic Multi-messenger Approach" number 2017X7X85K under the program PRIN 2017 funded by the Ministero dell'Istruzione, Universit$\grave{a}$ e della Ricerca (MIUR).
The work of XJB and PFY is supported by the National Natural Science Foundation of China under Grants No. U1738209.
\end{acknowledgments}

\bibliographystyle{utphys}
\bibliography{GCconvernsion}

\end{document}